\RequirePackage{fix-cm}
\documentclass[smallextended]{svjour3}       
\smartqed  
\usepackage{graphicx}
\usepackage{amsmath,amssymb}
\newcommand{\sdir}{\ensuremath{\rlap{\raisebox{.15ex}{$\mskip
6.5mu\scriptstyle+ $}}\supset}}
 \journalname{Ricerche di Matematica}
\begin{document}

\title{Invariant solutions of the supersymmetric version of a two-phase fluid flow system\thanks{This research was supported by a research grant from NSERC of Canada.}
}

\titlerunning{Invariant solutions of a supersymmetric two-phase fluid flow}        

\author{A. M. Grundland         \and
        A. J. Hariton 
}


\institute{A. M. Grundland \at
              Centre de Recherches Math{\'e}matiques, Universit{\'e} de Montr{\'e}al,\\
C. P. 6128, Succ.\ Centre-ville, Montr{\'e}al, (QC) H3C 3J7,
Canada\\ Universit\'{e} du Qu\'{e}bec, Trois-Rivi\`{e}res, CP500 (QC) G9A 5H7, Canada \\
              Tel.: (514) 343-6111 ext. 4741\\
              \email{grundlan@crm.umontreal.ca}           
           \and
           A. J. Hariton \at
              Centre de Recherches Math{\'e}matiques, Universit{\'e} de Montr{\'e}al,\\
C. P. 6128, Succ.\ Centre-ville, Montr{\'e}al, (QC) H3C 3J7,
Canada
}

\date{Received: date / Accepted: date}

\maketitle

\begin{abstract}
A supersymmetric extension of a two-phase fluid flow system is formulated. A superalgebra of Lie symmetries of the supersymmetric extension of this system is computed. The classification of the one-dimensional subalgebras of this superalgebra into 63 equivalence classes is performed. For some of the subalgebras, it is found that the invariants have a non-standard structure. For six selected subalgebras, the symmetry reduction method is used to find invariants, orbits of the subgroups and reduced systems. Through the solutions of the reduced systems, the most general solutions are  expressed in terms of arbitrary functions of one or two fermionic and one bosonic variables.
\end{abstract}

\section{Introduction}
\label{intro}
\noindent Among systems of partial differential equations (PDEs), quasilinear systems of PDEs with two independent variables have been the most studied. These systems describe, in particular, flows of compressible fluids. The basic method for solving these systems is the method of characteristics, which is described in detail in many books (see e.g. \cite{CourantHilbert,Mises,Lighthill,Whitham,Lamb,Jeffrey,Roz,Godunov,Jackiw,Majda,Peradzynski,Ovsiannikov,Zakharov} and references therein). The concept of differential invariants along an appropriate characteristic curve of a hyperbolic system of PDEs was first introduced by B. Riemann in 1853 \cite{Riemann1,Riemann2} in the context of two-wave superpositions determined by the Euler equations for an ideal compressible fluid flow in two independent variables
\begin{equation}
\dfrac{\partial u}{\partial t}+u\dfrac{\partial u}{\partial x}+\dfrac{1}{\rho}p'(\rho)\dfrac{\partial \rho}{\partial x}=0,\qquad\qquad \dfrac{\partial \rho}{\partial t}+u\dfrac{\partial \rho}{\partial x}+\rho \dfrac{\partial u}{\partial x}=0,\qquad\qquad \rho>0.
\label{eq1.1}
\end{equation}
Here, $u$ is the local velocity of the fluid, the pressure $p$ is assumed to be a differentiable function of the density $\rho$ and $p'=dp/d\rho$. The generalization of the plane wave solutions of the system (\ref{eq1.1}) were first introduced by S. D. Poisson \cite{Poisson} in 1808 for the equation describing compressible isothermal incompressible gas flows
\begin{equation}
\dfrac{\partial u}{\partial t}+u\dfrac{\partial u}{\partial x}=0.
\label{poisson1A}
\end{equation}
He constructed solutions of (\ref{poisson1A}) which can be interpreted as propagation waves, assuming that they take the form of implicit functions depending on one arbitrary function $F$ of one variable $s=(u+u_0)t-x$,
\begin{equation}
u=F((u+u_0)t-x),\qquad u_0\in\mathbb{R}.
\label{Riemann1Bw}
\end{equation}
Consequently, the derivatives of (\ref{Riemann1Bw}) with respect to $x$ and $t$ are
\begin{equation}
\dfrac{\partial u}{\partial x}=\dfrac{-F'}{1-F't}\hspace{1cm}\mbox{and}\hspace{1cm}\dfrac{\partial u}{\partial t}=\dfrac{uF'}{1-F't}
\label{introdnext1A}
\end{equation}
where $F'=dF/ds$. These derivatives are assumed to be different from zero. Poisson observed that the domain of existence of the implicit solution (\ref{Riemann1Bw}), in general, is limited since this solution of (\ref{poisson1A}) possesses the property of unbounded growth in the magnitude of the derivatives (\ref{introdnext1A}) when $t>0$. This would later be referred to in the literature as the gradient catastrophe \cite{CourantFriedrichs,Mises,Roz,Godunov}.
This idea was substantially generalized by B. Riemann who investigated the solvability of the Cauchy problem, the formulation and mathematical correctness of problems involving the propagation and superposition of waves described by the system (\ref{eq1.1}). Next, he constructed rank-2 solutions corresponding to the superposition of two waves propagating with local phase velocity $u=\pm(p'(\rho))^{1/2}$ when they propagate in parallel and opposite directions. This is known as the double waves problem. The main feature of the Riemann method is the introduction of new dependent variables (called the Riemann invariants). These invariants have the property that they preserve their values along appropriate characteristic curves of the initial system. This approach enables us to reduce the number of dependent variables for the wave superposition problem. Consequently, this method considerably simplifies the task of solving the initial system under consideration. In the case of the Euler equations (\ref{eq1.1}), this approach allows us to construct a scattering double wave solution (i.e. a rank-2 solution) written in terms of the Riemann invariants $r^i(t,x)$
\begin{equation}
u=k^{1/2}(r^1-r^2)+u_0,\qquad\qquad\rho=Ae^{r^1+r^2},\qquad\qquad p=kAe^{r^1+r^2}+p_0,
\end{equation}
where $u_0,p_0,k,A\in\mathbb{R}$ are constants of integration. Using the Riemann invariants $r^i(t,x)$, $i=1,2$, we can reduce the system (\ref{eq1.1}) to the system of PDEs
\begin{equation}
r^1_t+\left(k^{1/2}(r^1-r^2+1)+u_0\right)r^1_x=0,\qquad r^2_t+\left(k^{1/2}(r^1-r^2-1)+u_0\right)r^2_x=0.
\label{eq1.3}
\end{equation}
Here, each subscript means the partial derivative with respect to that variable (for example, $r^1_t$ means the partial derivative of $r^1$ with respect to $t$, and so on). Riemann also studied the asymptomatic behavior of the system (\ref{eq1.3}) for initial localized disturbances having initial data at $t=t_0$ with compact and disjoint supports corresponding to the considered waves. Consequently, he proved that even for sufficiently small initial differentiable initial data, after some finite time $T$, these waves could be separated again in such a way that the waves are of the same type as those imposed in the initial data (the so-called elastic superposition of two waves). Riemann noticed that the solution for hydrodynamic systems of the form (\ref{eq1.1}), even with arbitrary smooth initial data, usually cannot be continued indefinitely in time $t$. After a certain finite time $T$ the solutions blow up. More precisely, the first derivatives of the considered solution become unbounded after some finite time $T>0$. So, for times $t>T$, smooth solutions of the Cauchy problem do not exist. He showed that the gradient catastrophe can occur where the characteristics of the same family of the initial system intersect. Riemann extended the solution in some more generalized sense beyond the time $T$ of the blow up. Based on conservation laws for the mass, energy and momentum, Riemann \cite{Riemann1,Riemann2} and later H. Hugoniot \cite{Hugoniot} introduced the concept of weak solutions (noncontinuous) in the form of shock waves. Based on these conservation laws, they demonstrated some relations connecting the wave front velocity and parameters of the fluid state before and after that discontinuity.

\noindent The two-phase fluid flow described by the hyperbolic system \cite{Opanasenko1,Opanasenko2,Panov}
\begin{equation}
\begin{split}
& \dfrac{\partial \rho^1}{\partial t}+u\dfrac{\partial \rho^1}{\partial x}+\dfrac{\partial u}{\partial x}\rho^1=0,\qquad \dfrac{\partial \rho^2}{\partial t}+u\dfrac{\partial \rho^2}{\partial x}+\dfrac{\partial u}{\partial x}\rho^2=0,\\ &\qquad \left(\rho^1+\rho^2\right)\left(\dfrac{\partial u}{\partial t}+u\dfrac{\partial u}{\partial x}\right)+a^2\left(\dfrac{\partial \rho^1}{\partial x}+\dfrac{\partial \rho^2}{\partial x}\right)=0,
\end{split}
\label{eq1.4}
\end{equation}
where $\rho^1(t,x)$ and $\rho^2(t,x)$ are the densities of the two-phase fluid flow and the constant parameter $a$ can be set to one, possesses the same properties as the system (\ref{eq1.1}) and can also be written in terms of Riemann invariants
\begin{equation}
r^1_t+(r^1+r^2+1)r^1_x=0,\qquad r^2_t+(r^1+r^2-1)r^2_x=0,\qquad r^3_t+(r^1+r^2)r^3_x=0,\qquad
\end{equation}
using the change of variables
\begin{equation}
w=r^1+r^2,\qquad \rho^1+\rho^2=e^{r^1-r^2},\qquad \dfrac{\rho^2}{\rho^1}=r^3.
\end{equation}
The problem of the propagation and superposition of Riemann waves has been investigated by many authors (see e.g. \cite{John1,John2,Whitham,Jeffrey}). The task of finding and increasing the number of particular solutions is related to the group properties of hydrodynamic-type systems of the form (\ref{eq1.4}). The fact that Riemann simple waves (i.e. envelopes of center waves), later called self-similar solutions, are invariant under the Galilean similitude transformation, means that the system (\ref{eq1.4}) has a group character. The possibility of obtaining simple waves can be determined after an analysis of their group properties \cite{Ovsiannikov}. Several years later, it was realized \cite{Cornwell,DeWitt,Rogers1,Rogers2} that this and many other similar problems can be studied by modern methods developed in the theory of supersymmetric systems. This finally led to the development of new group analytic properties and algebraic methods in PDEs (i.e. analyzed from the Grassmann point of view). This resulted in, among other significant progress in the theory of supersymmetric (SUSY) extensions of PDEs, the classification of the subalgebras of superalgebras into conjugation classes under the action of the associated supergroup and the use of the symmetry reduction method to obtain several classes of invariant solutions of the supersymmetric system (see in particular \cite{Manin,Fatyga}). Recently, supersymmetric extensions of hydrodynamic-type systems has become a subject of intensive research (see e.g. \cite{Jackiw,Rogers1,Rogers2,EulerSUSY,Polytropic,GHhydroRiemann,Gaussian}). In view of the above, the objective of this paper is to construct a supersymmetric extension of the two-phase fluid flow equation (\ref{eq1.4}) and to study its supersymmetry properties. This approach allows us to reduce the extended SUSY system related to group properties and to find new particular classes of invariant solutions. 

\noindent This paper is organized as follows. In section 2, we recall some basic facts, notions and definitions and formulate a supersymmetric extension of the hydrodynamic equations (\ref{eq1.4}) through a superspace and superfield formalism. In section 3, we present a superalgebra of infinitesimal symmetries of our constructed supersymmetric system, and classify the one-dimensional subalgebras of this superalgebra into conjugation classes under the action of the associated supergroup. The subalgebras are classified in such a way that each representative subalgebra corresponding to a different  conjugation class generates a different type of solution. In section 4, based on this analysis, we use the symmetry reduction method to construct several new classes of invariant solutions of the supersymmetric two-phase fluid flow equations. Finally, section 5 contains concluding remarks and a description of possible future developments. In the Appendix, we present the list of the 63 one-dimensional Lie subalgebras associated with the supersymmetric system.

\section{Supersymmetric extension}
\label{sec:2}

In order to supersymmetrize the two-phase fluid flow system (\ref{eq1.4}), it is necessary to review the following concepts in accordance with \cite{Cornwell,DeWitt,Rogers1,Rogers2}. Grassmann variables are elements of a Grassmann algebra $\Lambda$ involving a finite number of Grassmann generators $\zeta_1,\zeta_2,\ldots,\zeta_k$ which obey the rules
\begin{equation}
\zeta_i\zeta_j=-\zeta_j\zeta_i\mbox{ if }i\neq j,\qquad
\zeta_i^2=0\mbox{ for all }i.
\label{grasscond}
\end{equation}
The Grassmann algebra can be decomposed into even and odd parts: $\Lambda=\Lambda_{even}+\Lambda_{odd}$, where $\Lambda_{even}$ consists of all terms involving the product of an even number of generators $1,\zeta_1\zeta_2,\zeta_1\zeta_3,\ldots,\zeta_1\zeta_2\zeta_3\zeta_4,\ldots$, while $\Lambda_{odd}$ consists of all terms involving the product of an odd number of generators $\zeta_1,\zeta_2,\zeta_3,\ldots,\zeta_1\zeta_2\zeta_3,\zeta_1\zeta_2\zeta_4,\ldots$. A Grassmann variable $k$ is called even (or {\bf bosonic}) if it is a linear combination of terms involving an even number of generators, while it is called odd (or {\bf fermionic}) if it is a linear combination of terms involving an odd number of generators.\\\\
Let us now construct a Grassmann-valued extension of the two-phase fluid flow system (\ref{eq1.4}).
The space of (bosonic Grassmann-valued) independent variables, $\{(x,t)\}$, is extended to a four-dimensional superspace $\{(x,t,\theta_1,\theta_2)\}$ involving two fermionic Grassmann-valued variables $\theta_1$ and $\theta_2$. Also, the bosonic functions $\rho^1(x,t)$, $\rho^2(x,t)$ and $u(x,t)$ are generalized to the bosonic-valued superfields $A(x,t,\theta_1,\theta_2)$, $B(x,t,\theta_1,\theta_2)$ and $\Phi(x,t,\theta_1,\theta_2)$ respectively defined as
\begin{equation}
\begin{split}
&A(x,t,\theta_1,\theta_2)=a(x,t)+\theta_1b(x,t)+\theta_2c(x,t)+\theta_1\theta_2\rho^1(x,t),\\
&B(x,t,\theta_1,\theta_2)=f(x,t)+\theta_1g(x,t)+\theta_2h(x,t)+\theta_1\theta_2\rho^2(x,t),\\
&\Phi(x,t,\theta_1,\theta_2)=m(x,t)+\theta_1q(x,t)+\theta_2s(x,t)+\theta_1\theta_2u(x,t),
\end{split}
\label{superfield}
\end{equation}
where the functions $a$, $f$ and $m$ are bosonic-valued fields of $(x,t)$, while $b$, $c$, $g$, $h$, $q$ and $s$ are fermionic-valued fields of $(x,t)$. The superfields (\ref{superfield}) take values in a Grassmann ring. We construct our extension in such a way that it is invariant under the supersymmetry transformations
\begin{equation}
x\longrightarrow x-\underline{\eta_1}\theta_1,\hspace{5mm}\theta_1\longrightarrow \theta_1+\underline{\eta_1},
\label{trQ1}
\end{equation}
and
\begin{equation}
t\longrightarrow t-\underline{\eta_2}\theta_2,\hspace{5mm}\theta_2\longrightarrow \theta_2+\underline{\eta_2},
\label{trQ2}
\end{equation}
where $\underline{\eta_1}$ and $\underline{\eta_2}$ are odd-valued parameters. Throughout this paper, we use the convention that underlined constants are fermionic-valued. The transformations (\ref{trQ1}) and (\ref{trQ2}) are generated by the infinitesimal supersymmetry generators
\begin{equation}
Q_1=\partial_{\theta_1}-\theta_1\partial_{x}\hspace{1cm}\mbox{and}\hspace{1cm}Q_2=\partial_{\theta_2}-\theta_2\partial_{y},
\label{supersymmetry}
\end{equation}
respectively. These generators satisfy the anticommutation relations
\begin{equation}
\{Q_1,Q_1\}=-2\partial_x,\hspace{1cm}\{Q_2,Q_2\}=-2\partial_y,\hspace{1cm}\{Q_1,Q_2\}=0.
\label{anticommutators}
\end{equation}
To make the superfield model invariant under the transformations generated by $Q_1$ and $Q_2$, we construct the equation in terms of the following covariant derivatives:
\begin{equation}
D_1=\partial_{\theta_1}+\theta_1\partial_{x}\hspace{1cm}\mbox{and}\hspace{1cm}D_2=\partial_{\theta_2}+\theta_2\partial_{y}.
\label{covariant}
\end{equation}
Each of these covariant derivative operators, $D_i$, $i=1,2$, anticommutes with every supersymmetry operator $Q_j$, so we obtain the following properties:
\begin{equation}
\begin{split}
& D_1^2=2\partial_x,\hspace{1cm}D_2^2=2\partial_y,\hspace{1cm}\{D_1,D_2\}=0,\hspace{1cm}\{D_1,Q_1\}=0,\hspace{1cm}\\ & \{D_1,Q_2\}=0,\hspace{1cm}\{D_2,Q_1\}=0,\hspace{1cm}\{D_2,Q_2\}=0.
\end{split}
\label{covprop}
\end{equation}
Combining different covariant derivatives $D_1^m$ and $D_2^n$ of the superfield $\Phi$ of various orders, where $m$ and $n$ are positive integers, we obtain the most general form of the supersymmetric extension the system (\ref{eq1.4}). Since this expression is very involved, we instead present the following sub-case as our supersymmetric extension of the two-phase system, and will refer to it as such. We obtain the system
\begin{equation}
D_2^2A-D_1^3D_2A\Phi-(D_1D_2A)(D_1^2\Phi)=0,
\label{eqmotion1A}
\end{equation}
\begin{equation}
D_2^2B-D_1^3D_2B\Phi-(D_1D_2B)(D_1^2\Phi)=0,
\label{eqmotion2A}
\end{equation}
\begin{equation}
\begin{split}
& A(D_1D_2^3\Phi)+(D_1A)(D_2\Phi)(D_1^3D_2\Phi)+B(D_1D_2^3\Phi)+(D_1B)(D_2\Phi)(D_1^3D_2\Phi)\\ & +D_1^2A+D_1^2B=0.
\end{split}
\label{eqmotion3A}
\end{equation}

In terms of derivatives with respect to $x$, $t$, $\theta_1$ and $\theta_2$, the system formed of equations (\ref{eqmotion1A}), (\ref{eqmotion2A}) and (\ref{eqmotion3A}) can be written in the form
\begin{equation}
\begin{split}
& A_t-\theta_1\theta_2A_x\Phi_{xt}+\theta_2A_x\Phi_{t\theta_1}-\theta_1A_x\Phi_{x\theta_2}+A_x\Phi_{\theta_1\theta_2}-\theta_1\theta_2A_{xt}\Phi_x+\theta_2A_{t\theta_1}\Phi_x\\ & -\theta_1A_{x\theta_2}\Phi_x+A_{\theta_1\theta_2}\Phi_x=0,
\end{split}
\label{eqmotion1B}
\end{equation}
\begin{equation}
\begin{split}
& B_t-\theta_1\theta_2B_x\Phi_{xt}+\theta_2B_x\Phi_{t\theta_1}-\theta_1B_x\Phi_{x\theta_2}+B_x\Phi_{\theta_1\theta_2}-\theta_1\theta_2B_{xt}\Phi_x+\theta_2B_{t\theta_1}\Phi_x\\ & -\theta_1B_{x\theta_2}\Phi_x+B_{\theta_1\theta_2}\Phi_x=0,
\end{split}
\label{eqmotion2B}
\end{equation}
\begin{equation}
\begin{split}
& \theta_1\theta_2A_x\Phi_{tt}-\theta_2A_{\theta_1}\Phi_{tt}+\theta_1A_x\Phi_{t\theta_2}+A_{\theta_1}\Phi_{t\theta_2}-\theta_1\theta_2A_{xt}\Phi_{\theta_1\theta_2}\Phi_x-\theta_1\theta_2A_{t\theta_1}\Phi_{x\theta_2}\Phi_x\\ & +\theta_2A_{t\theta_1}\Phi_{\theta_1\theta_2}\Phi_x+\theta_1\theta_2A_{x\theta_2}\Phi_{t\theta_1}\Phi_x-\theta_1A_{x\theta_2}\Phi_{\theta_1\theta_2}\Phi_x-\theta_1\theta_2A_{\theta_1\theta_2}\Phi_{xt}\Phi_x\\ & +\theta_2A_{\theta_1\theta_2}\Phi_{t\theta_1}\Phi_x-\theta_1A_{\theta_1\theta_2}\Phi_{x\theta_2}\Phi_x+A_{\theta_1\theta_2}\Phi_{\theta_1\theta_2}\Phi_x+\theta_1\theta_2B_x\Phi_{tt}-\theta_2B_{\theta_1}\Phi_{tt}\\ & +\theta_1B_x\Phi_{t\theta_2}+B_{\theta_1}\Phi_{t\theta_2}-\theta_1\theta_2B_{xt}\Phi_{\theta_1\theta_2}\Phi_x-\theta_1\theta_2B_{t\theta_1}\Phi_{x\theta_2}\Phi_x+\theta_2B_{t\theta_1}\Phi_{\theta_1\theta_2}\Phi_x\\ & +\theta_1\theta_2B_{x\theta_2}\Phi_{t\theta_1}\Phi_x-\theta_1B_{x\theta_2}\Phi_{\theta_1\theta_2}\Phi_x-\theta_1\theta_2B_{\theta_1\theta_2}\Phi_{xt}\Phi_x+\theta_2B_{\theta_1\theta_2}\Phi_{t\theta_1}\Phi_x\\ & -\theta_1B_{\theta_1\theta_2}\Phi_{x\theta_2}\Phi_x+B_{\theta_1\theta_2}\Phi_{\theta_1\theta_2}\Phi_x+A_x+B_x=0.
\end{split}
\label{eqmotion3B}
\end{equation}
In what follows, we will refer to the system formed of equations (\ref{eqmotion1B}), (\ref{eqmotion2B}) and (\ref{eqmotion3B}) as the supersymmetric two-phase system (SUSY two-phase system). If one considers the limit case where the fermionic variables $\theta_1$ and $\theta_2$ approach zero, the system (\ref{eqmotion1B}), (\ref{eqmotion2B}) and (\ref{eqmotion3B}) reduces to the classical system (\ref{eq1.4}), which can also be written in terms of Riemann invariants.\\\\
The partial derivatives satisfy the generalized Leibniz rule
\begin{equation}
\partial_{\theta_i}(fg)=(\partial_{\theta_i}f)g+(-1)^{\mbox{deg}(f)}f(\partial_{\theta_i}g),
\end{equation}
if $\theta_i$ is a fermionic variable and we define
\begin{equation}
\mbox{deg}(f)=\begin{cases}
               0 & \mbox{ if } f \mbox{ is even},\\
               1 & \mbox{ if } f \mbox{ is odd}.
              \end{cases}
\end{equation}
The partial derivatives with respect to the odd coordinates satisfy $\partial_{\theta_i}(\theta_j)=\delta_j^i$, where the indices $i$ and $j$ each stand for 1 or 2 and $\delta_j^i$ is the Kronecker delta function. The differential operators $\partial_{\theta_1}$, $\partial_{\theta_2}$, $Q_1$, $Q_2$, $D_1$ and $D_2$ change the parity of a bosonic function to that of a fermionic function and vice versa.\\\\
When dealing with higher-order derivatives, the symbol $f_{x_1x_2x_3\ldots x_{k-1}x_k}$ denotes the derivative $\partial_{x_k}\partial_{x_{k-1}}\ldots\partial_{x_3}\partial_{x_2}\partial_{x_1}(f)$ where the order must be preserved for the sake of the consistency of subsequent operators.
Throughout this paper, we use the convention that if $f(g(x))$ is a composite function, then the chain rule preserves the order
\begin{equation}
\dfrac{\partial f}{\partial x}=\dfrac{\partial g}{\partial x}\cdot\dfrac{\partial f}{\partial g}.
\label{composite}
\end{equation}
The interchange of mixed derivatives with proper respect for the ordering of odd variables is assumed throughout. For a review of recent developments in this subject see e.g. Freed \cite{Freed} and Varadarajan \cite{Varadarajan}.

\section{Lie superalgebra and classification of its subalgebras}
\label{sec:3}

A symmetry supergroup $G$ of a supersymmetric system is a local supergroup of transformations acting on the Cartesian product of submanifolds ${\mathcal X}\times{\mathcal U}$, where ${\mathcal X}$ is the space of independent variables $\{(x,t,\theta_1,\theta_2)\}$ and ${\mathcal U}$ is the space of dependent superfields $\{(A,B,\Phi)\}$. For the purpose of finding symmetries of the SUSY two-phase system, we make use of the theory described in the book by Olver \cite{Olver} to determine superalgebras of infinitesimal symmetries.\\\\
In order to find a Lie point superalgebra of infinitesimal symmetries, we look for an infinitesimal vector field of the form
\begin{equation}
\begin{split}
{\mathbf v}=&\xi(x,t,\theta_1,\theta_2,A,B,\Phi)\partial_x+\tau(x,t,\theta_1,\theta_2,A,B,\Phi)\partial_t+\rho_1(x,t,\theta_1,\theta_2,A,B,\Phi)\partial_{\theta_1}\\ & +\rho_2(x,t,\theta_1,\theta_2,A,B,\Phi)\partial_{\theta_2}+\Lambda(x,t,\theta_1,\theta_2,A,B,\Phi)\partial_A\\ & +\Psi(x,t,\theta_1,\theta_2,A,B,\Phi)\partial_B+\Omega(x,t,\theta_1,\theta_2,A,B,\Phi)\partial_{\Phi},
\label{vectorfield}
\end{split}
\end{equation}
where $\xi$, $\tau$, $\rho_1$, $\rho_2$, $\Lambda$, $\Psi$ and $\Omega$ are functions of $(x,t,\theta_1,\theta_2,A,B,\Phi)$. 
The following infinitesimal point transformations were found to be symmetry generators of the SUSY two-phase system (\ref{eqmotion1B}), (\ref{eqmotion2B}) and (\ref{eqmotion3B}):
\begin{equation}
\begin{split}
& P_1=\partial_x,\hspace{1cm}P_2=\partial_t,\hspace{1cm}M_1=2x\partial_x+2t\partial_t+\theta_1\partial_{\theta_1}+\theta_2\partial_{\theta_2}+2\Phi\partial_{\Phi},\\ & M_2=A\partial_A+B\partial_B,\hspace{1cm} Q_1=\partial_{\theta_1}-\theta_1\partial_{x},\hspace{1cm} Q_2=\partial_{\theta_2}-\theta_2\partial_{t}.
\end{split}
\label{symmetries}
\end{equation}
These six generators span a Lie superalgebra ${\mathcal L}$ of infinitesimal symmetries of the SUSY two-phase system. Here, $P_1$ and $P_2$ generate translations in the $x$ and $t$ directions respectively. The vector field $M_1$ corresponds to a dilation involving both bosonic and fermionic variables as well as the superfield $\Phi$, while $M_2$ corresponds to a dilation in the superfields $A$ and $B$. Finally, the fermionic vector fields $Q_1$ and $Q_2$ are simply the supersymmetry transformations identified in (\ref{supersymmetry}). The supercommutation relations involving the generators of the superalgebra ${\mathcal L}$ are listed in Table 1. 

\begin{table}[htbp]
\caption{Supercommutation table for the Lie point superalgebra ${\mathcal L}$ generated by the vector fields (\ref{symmetries}). Here, for each pair of generators $X$ and $Y$, we calculate either the commutator $[X,Y]=XY-YX$ if either $X$ or $Y$ is bosonic, or the anticommutator $\{X,Y\}=XY+YX$ if both $X$ and $Y$ are fermionic.}
\begin{center}
\begin{tabular}{|c||c|c|c|c|c|c|}\hline
 & $\mathbf{M_2}$ & $\mathbf{M_1}$ & $\mathbf{P_1}$ & $\mathbf{Q_1}$ & $\mathbf{P_2}$ & $\mathbf{Q_2}$ \\\hline\hline
$\mathbf{M_2}$ & $0$ & $0$ & $0$ & $0$ & $0$ & $0$ \\\hline
$\mathbf{M_1}$ & $0$ & $0$ & $-2P_1$ & $-Q_1$ & $-2P_2$ & $-Q_2$ \\\hline
$\mathbf{P_1}$ & $0$ & $2P_1$ & $0$ & $0$ & $0$ & $0$ \\\hline
$\mathbf{Q_1}$ & $0$ & $Q_1$ & $0$ & $-2P_1$ & $0$ & $0$ \\\hline
$\mathbf{P_2}$ & $0$ & $2P_2$ & $0$ & $0$ & $0$ & $0$ \\\hline
$\mathbf{Q_2}$ & $0$ & $Q_2$ & $0$ & $0$ & $0$ & $-2P_2$ \\\hline
\end{tabular}
\end{center}
\end{table}
We note that dilations and translations are also present among the symmetries of the classcial two-phase fluid flow equations (\ref{eq1.4}) \cite{Opanasenko1,Opanasenko2}. Also, Riemann simple waves of the compressible fluid equations are envelopes of centered waves (called self-similar \cite{Roz,Sidorov,Ovsiannikov}), which are invariant under scaling transformations involving bosonic and fermionic variables in the four-dimensional Grassmann superspace. This means that equations (\ref{eqmotion1B}), (\ref{eqmotion2B}) and (\ref{eqmotion3B}) have the group character.

\noindent The Lie superalgebra ${\mathcal L}$ can be decomposed into the following combination of semidirect and direct sums:
\begin{equation}
{\mathcal L}=\{M_2\}\oplus\{\{M_1\}\sdir\{\{P_1,Q_1\}\oplus\{P_2,Q_2\}\}\}
\label{decomposed}
\end{equation}

We proceed to classify the one-dimensional Lie subalgebras of the superalgebra ${\mathcal L}$ generated by (\ref{symmetries}) into conjugacy classes under the action of the Lie supergroup $G=\mbox{exp}({\mathcal L})$ generated by ${\mathcal L}$. We construct our list of representative subalgebras in such a way that each one-dimensional subalgebra of ${\mathcal L}$ is conjugate to one and only one element of the list. Such a classification is useful because subalgebras that are conjugate to each other lead to invariant solutions that are equivalent in the sense that one can be transformed to the other by a suitable symmetry. Therefore, it is not necessary to perform symmetry reductions on two different subalgebras that are conjugate to each other.\\\\
In order to classify the Lie superalgebra ${\mathcal L}$ given in (\ref{decomposed}) we make use of the procedures given in \cite{Winternitz}. In what follows, $\alpha$, $r$, $k$ and $\ell$ are bosonic constants, $\underline{\mu}$, $\underline{\nu}$, $\underline{\eta}$, $\underline{\lambda}$, $\underline{\rho}$ and $\underline{\sigma}$ are fermionic constants, and $\varepsilon=\pm 1$. We begin by considering the subalgebra ${\mathcal A}=\{P_1,Q_1\}$. Consider a general element of ${\mathcal A}$ which can be written as the linear combination $X=\alpha P_1+\underline{\mu}Q_1$ and examine how this element changes under the action of the one-parameter group generated by the generator: $Y=rP_1+\underline{\eta}Q_1$. This action is performed through the Baker-Campbell-Hausdorff formula
\begin{equation}
X\longrightarrow \mbox{Ad}_{\mbox{exp}(Y)}X=X+[Y,X]+\frac{1}{2\!}[Y,[Y,X]]+\ldots+\frac{1}{3\!}[Y,[Y,[Y,X]]]+\ldots
\label{BCHformula}
\end{equation}
We obtain
\begin{equation}
\begin{split}
[Y,X]&=[rP_1+\underline{\eta}Q_1,\alpha P_1+\underline{\mu}Q_1]\\
     &=[rP_1,\alpha P_1]+[rP_1,\underline{\mu}Q_1]+[\underline{\eta}Q_1,\alpha P_1]+[\underline{\eta}Q_1,\underline{\mu}Q_1]\\
     &=2\underline{\eta}\underline{\mu}P_1.
\end{split}
\end{equation}
Hence
\begin{equation}
[Y[Y,X]]=0.
\end{equation}
So we have
\begin{equation}
\{\alpha P_1+\underline{\mu}Q_1\}\longrightarrow \{(\alpha+2\underline{\eta}\underline{\mu}) P_1+\underline{\mu}Q_1\}.
\end{equation}
Therefore, aside from a change in the $P_1$ coefficient, each element of the form $\{\alpha P_1+\underline{\mu}Q_1\}$ is conjugate only to itself. This gives us the subalgebras
\begin{equation}
\begin{split}
&{\mathcal A}_1=\{P_1\}, \hspace{5mm} {\mathcal A}_2=\{\underline{\mu}Q_1\}, \hspace{5mm}
{\mathcal A}_3=\{P_1+\underline{\mu}Q_1\}.
\end{split}
\label{list1}
\end{equation}
An analogous classification is performed for the subalgebra ${\mathcal B}=\{P_2,Q_2\}$, from where we obtain the subalgebras
\begin{equation}
\begin{split}
&{\mathcal B}_1=\{P_2\}, \hspace{5mm} {\mathcal B}_2=\{\underline{\mu}Q_2\}, \hspace{5mm}
{\mathcal B}_3=\{P_2+\underline{\mu}Q_2\}.
\end{split}
\label{list2}
\end{equation}
\\
The next step is to classify the direct sum of the algebras ${\mathcal A}$ and ${\mathcal B}$, that is, to classify
\begin{equation}
{\mathcal C}={\mathcal A}\oplus {\mathcal B}=\{P_1,Q_1\}\oplus \{P_2,Q_2\},
\end{equation}
using the Goursat method of subalgebra classification \cite{Winternitz,Goursat1}. Each non-twisted subalgebra of ${\mathcal C}$ is constructed by selecting one subalgebra of ${\mathcal A}$ and finding its direct sum with a subalgebra of ${\mathcal B}$. The non-twisted one-dimensional subalgebras of ${\mathcal C}$ are the combined subalgebras of ${\mathcal A}$ and ${\mathcal B}$ listed in (\ref{list1}) and (\ref{list2}). The twisted subalgebras of ${\mathcal C}$ are formed as follows. If $A\in {\mathcal A}$ and $B\in {\mathcal B}$, then $A$ and $B$ can be twisted together if there exists a homomorphism from $A$ to $B$, say $\tau(A)=B$. The twisted subalgebra is then obtained by taking $\{A+\tau(A)\}$. The one-dimensional subalgebras of ${\mathcal C}$ are then
\begin{equation}
\begin{split}
&{\mathcal C}_1=\{P_1\}, \hspace{5mm} {\mathcal C}_2=\{\underline{\mu}Q_1\}, \hspace{5mm}
{\mathcal C}_3=\{P_1+\underline{\mu}Q_1\}, \hspace{5mm} {\mathcal C}_4=\{P_2\}, \hspace{5mm} {\mathcal C}_5=\{\underline{\mu}Q_2\}, \\ &
{\mathcal C}_6=\{P_2+\underline{\mu}Q_2\}, \hspace{5mm} {\mathcal C}_7=\{P_1+aP_2\}, \hspace{5mm} {\mathcal C}_8=\{P_1+\underline{\mu}Q_2\}, \hspace{5mm} {\mathcal C}_9=\{P_2+\underline{\mu}Q_1\}, \\ & {\mathcal C}_{10}=\{P_1+aP_2+\underline{\mu}Q_1\}, \hspace{5mm}
{\mathcal C}_{11}=\{P_1+aP_2+\underline{\mu}Q_2\}, \hspace{5mm} {\mathcal C}_{12}=\{\underline{\mu}Q_1+\underline{\nu}Q_2\}, \\ &
{\mathcal C}_{13}=\{P_1+\underline{\mu}Q_1+\underline{\nu}Q_2\}, \hspace{5mm} {\mathcal C}_{14}=\{P_2+\underline{\mu}Q_1+\underline{\nu}Q_2\}, \\ &
{\mathcal C}_{15}=\{P_1+aP_2+\underline{\mu}Q_1+\underline{\nu}Q_2\}.
\end{split}
\label{list3}
\end{equation}

\noindent Next, we classify the one-dimensional subalgebras of the semi-direct sum 
\begin{equation}
{\mathcal G}=\{M_1\}\sdir{\mathcal C}=\{M_1\}\sdir\{\{P_1,Q_1\}\oplus\{P_2,Q_2\}\}
\end{equation}
using the method of splitting and non-splitting subalgebras \cite{Clarkson,Winternitz}. The splitting subalgebras of ${\mathcal G}$ are formed by combining the dilation $\{M_1\}$ or the trivial element $\{0\}$ with each of the subalgebras of ${\mathcal C}$ in a semidirect sum of the form $F\sdir N$, where $F=\{M_1\}$ or $F=\{0\}$ and $N$ is a subalgebra of the classification of ${\mathcal C}$.  The splitting one-dimensional subalgebras of ${\mathcal G}$ consist of the subalgebras listed in (\ref{list3}) together with the subalgebra $\{M_1\}$. For non-splitting subalgebras, we consider spaces of the form
\begin{equation}
V=\{M_1+\sum\limits_{i=1}^{s}c_iZ_i\},
\end{equation}
where the $Z_i$ form a basis of the non-twisted one-dimensional subalgebra ${\mathcal C}$. The resulting possibilities are further classified by observing which are conjugate to each other under the action of the complete group generated by ${\mathcal G}$. This analysis provides us with the following classification:

\begin{equation}
\begin{split}
&{\mathcal G}_1=\{P_1\}, \hspace{5mm} {\mathcal G}_2=\{P_2\}, \hspace{5mm}
{\mathcal G}_3=\{P_1+aP_2\}, \hspace{5mm} {\mathcal G}_4=\{\underline{\mu}Q_1\}, \hspace{5mm} {\mathcal G}_5=\{\underline{\mu}Q_2\}, \\ &
{\mathcal G}_6=\{P_1+\underline{\mu}Q_1\}, \hspace{5mm} {\mathcal G}_7=\{P_1+\underline{\mu}Q_2\}, \hspace{5mm} {\mathcal G}_8=\{P_2+\underline{\mu}Q_1\}, \hspace{5mm} {\mathcal G}_9=\{P_2+\underline{\mu}Q_2\}, \\ & {\mathcal G}_{10}=\{P_1+aP_2+\underline{\mu}Q_1\}, \hspace{5mm}
{\mathcal G}_{11}=\{P_1+aP_2+\underline{\mu}Q_2\}, \hspace{5mm} {\mathcal G}_{12}=\{\underline{\mu}Q_1+\underline{\nu}Q_2\}, \\ &
{\mathcal G}_{13}=\{P_1+\underline{\mu}Q_1+\underline{\nu}Q_2\}, \hspace{5mm} {\mathcal G}_{14}=\{P_2+\underline{\mu}Q_1+\underline{\nu}Q_2\}, \\ & {\mathcal G}_{15}=\{P_1+aP_2+\underline{\mu}Q_1+\underline{\nu}Q_2\}, \hspace{5mm} {\mathcal G}_{16}=\{M_1\}, \hspace{5mm} 
{\mathcal G}_{17}=\{M_1+\varepsilon P_1\}, \\ & {\mathcal G}_{18}=\{M_1+\varepsilon P_2\},  \hspace{5mm}
{\mathcal G}_{19}=\{M_1+\varepsilon P_1+aP_2\}, \hspace{5mm} {\mathcal G}_{20}=\{M_1+\underline{\mu}Q_1\},  \\ &
{\mathcal G}_{21}=\{M_1+\underline{\mu}Q_2\}, \hspace{5mm} {\mathcal G}_{22}=\{M_1+\varepsilon P_1+\underline{\mu}Q_1\},  \hspace{5mm}
{\mathcal G}_{23}=\{M_1+\varepsilon P_1+\underline{\mu}Q_2\}, \\ & {\mathcal G}_{24}=\{M_1+\varepsilon P_2+\underline{\mu}Q_1\},  \hspace{5mm}
{\mathcal G}_{25}=\{M_1+\varepsilon P_2+\underline{\mu}Q_2\}, \\ & {\mathcal G}_{26}=\{M_1+\varepsilon P_1+aP_2+\underline{\mu}Q_1\},  \hspace{5mm}
{\mathcal G}_{27}=\{M_1+\varepsilon P_1+aP_2+\underline{\mu}Q_2\}, \\ & {\mathcal G}_{28}=\{M_1+\underline{\mu}Q_1+\underline{\nu}Q_2\},  \hspace{5mm}
{\mathcal G}_{29}=\{M_1+\varepsilon P_1+\underline{\mu}Q_1+\underline{\nu}Q_2\}, \\ & {\mathcal G}_{30}=\{M_1+\varepsilon P_2+\underline{\mu}Q_1+\underline{\nu}Q_2\}, \hspace{5mm} {\mathcal G}_{31}=\{M_1+\varepsilon P_1+aP_2+\underline{\mu}Q_1+\underline{\nu}Q_2\}.
\end{split}
\label{list4}
\end{equation}

\ \\
Finally, we classify the complete superalgebra ${\mathcal L}=\{M_2\}\oplus{\mathcal G}$, again using the Goursat method, obtaining a list of 63 non-equivalent conjugacy classes of one-dimensional subalgebras of the superalgebra ${\mathcal L}$ associated with the supersymmetric system (\ref{eqmotion1B}), (\ref{eqmotion2B}) and (\ref{eqmotion3B}). This list is presented in the Appendix.

\section{Invariant solutions of the supersymmetric two-phase flow}
\label{sec:4}

Each subalgebra given in the Appendix can be used to perform a symmetry reduction of the supersymmetric two-phase system composed of equations (\ref{eqmotion1B}), (\ref{eqmotion2B}) and (\ref{eqmotion3B}) which, in most cases, allows us to determine invariant solutions of the supersymmetric two-phase system. Once an invariant solution of the system is known, new solutions can be found by acting on the given solution with the supergroup of symmetries. Since both the system and the list of subalgebras are very involved, we do not consider all possible cases. Instead we present certain interesting examples of nontrivial solutions which illustrate the symmetry reduction method. In each case, we begin by constructing a complete set of invariants (functions which are preserved by a symmetry subgroup action). Next, we find the group orbits of the corresponding subgroups as well as the associated reduced systems of equations. Each reduced system can be solved in order to construct an invariant solution of the supersymmetric two-phase system with equations (\ref{eqmotion1B}), (\ref{eqmotion2B}) and (\ref{eqmotion3B}). It should be noted that, as has been observed for other similar supersymmetric extensions of SUSY hydrodynamic-type systems \cite{Polytropic,GHhydroRiemann,Gaussian}, some of the subalgebras listed in the Appendix have a non-standard invariant structure in the sense that they do not reduce the system to ordinary differential equations in the usual sense. These are the 7 subalgebras ${\mathcal L}_4$, ${\mathcal L}_8$, ${\mathcal L}_{12}$, ${\mathcal L}_{32}$, ${\mathcal L}_{36}$, ${\mathcal L}_{40}$ and ${\mathcal L}_{44}$. This leaves 56 subalgebras that lead to standard symmetry reductions, of which we illustrate several examples. In what follows, we use the following notation for arbitrary constants: $C_1$, $\underline{C_2}$,  $\underline{C_3}$, $C_4$, $C_5$, $\underline{C_6}$, $\underline{C_7}$, $C_8$, $A_0$, $\underline{A_1}$, $\underline{A_2}$, $A_3$, $B_1$, $\underline{B_2}$, $\underline{B_3}$, $B_4$, $F_1$, $\underline{F_2}$, $\underline{F_3}$, $F_4$, $\underline{\Phi_1}$, $\underline{\Phi_2}$, $\Phi_3$.\\

\noindent{\bf 1.} For the subalgebra ${\mathcal L}_1=\{\partial_x\}$, the invariants are $t$, $\theta_1$, $\theta_2$, $A$, $B$ and $\Phi$, which leads to the group orbit $A=A(t,\theta_1,\theta_2)$, $B=B(t,\theta_1,\theta_2)$, $\Phi=\Phi(t,\theta_1,\theta_2)$. Substituting into the SUSY two-phase equations (\ref{eqmotion1B}), (\ref{eqmotion2B}) and (\ref{eqmotion3B}), we obtain three different classes of exact solutions. \\

\noindent Equation (\ref{eqmotion1B}) becomes $A_t=0$, so $A$ is a function of $\theta_1$ and $\theta_2$ only:
\begin{equation}
A=K_1\theta_1\theta_2+\underline{K_2}\theta_2+\underline{K_3}\theta_1+K_4.
\label{thefieldA}
\end{equation}
Equation (\ref{eqmotion2B}) becomes $B_t=0$, so $B$ is a function of $\theta_1$ and $\theta_2$ only:
\begin{equation}
B=K_5\theta_1\theta_2+\underline{K_6}\theta_2+\underline{K_7}\theta_1+K_8.
\label{thefieldB}
\end{equation}
Equation (\ref{eqmotion3B}) becomes
\begin{equation}
-\theta_2A_{\theta_1}\Phi_{tt}+A_{\theta_1}\Phi_{t\theta_2}-\theta_2B_{\theta_1}\Phi_{tt}+B_{\theta_1}\Phi_{t\theta_2}=0,
\label{thefieldsAandB}
\end{equation}
and so, substituting equations (\ref{thefieldA}) and (\ref{thefieldB}) into (\ref{thefieldsAandB}), we obtain
\begin{equation}
\theta_2\left(\underline{K_3}\Phi_{tt}+K_1\Phi_{t\theta_2}+\underline{K_7}\Phi_{tt}+K_5\Phi_{t\theta_2}\right)-\left(\underline{K_3}+\underline{K_7}\right)\Phi_{t\theta_2}=0,
\end{equation}
from where we obtain the two equations
\begin{equation}
\left(\underline{K_3}+\underline{K_7}\right)\Phi_{t\theta_2}=0
\label{manycases1A3}
\end{equation}
and
\begin{equation}
\left(\underline{K_3}+\underline{K_7}\right)\Phi_{tt}+\left(K_1+K_5\right)\Phi_{t\theta_2}=0
\label{manycases1A4}
\end{equation}
We have three different cases.\\

\noindent In the case where $\underline{K_3}+\underline{K_7}=0$ and $K_1+K_5=0$, we obtain, after re-labeling $K_1=C_1$, $\underline{K_2}=\underline{C_2}$, $\underline{K_3}=\underline{C_3}$, $K_4=C_4$, $\underline{K_6}=\underline{C_5}$, $K_8=C_6$, the solution
\begin{equation}
\begin{split}
&A(\theta_1,\theta_2)=C_1\theta_1\theta_2+\underline{C_2}\theta_2+\underline{C_3}\theta_1+C_4\\
&B(\theta_1,\theta_2)=-C_1\theta_1\theta_2+\underline{C_5}\theta_2-\underline{C_3}\theta_1+C_6\\
&\Phi=\Phi(t,\theta_1,\theta_2)
\end{split}
\label{solutionG1}
\end{equation}

\noindent In the case where $\underline{K_3}+\underline{K_7}=0$ but $K_1+K_5\neq 0$, we have $\Phi_{t\theta_2}=0$. Since $\Phi$ is of the form
\begin{equation}
\Phi(t,\theta_1,\theta_2)=\alpha(t)\theta_1\theta_2+\beta(t)\theta_2+\gamma(t)\theta_1+\delta(t),
\label{formofphib612}
\end{equation}
where $\alpha$, $\beta$, $\gamma$ and $\delta$ are arbitrary functions of $t$, we obtain $-\alpha_t\theta_1-\beta_t=0$. Therefore, $\alpha_t=\beta_t=0$, so that $\alpha$ and $\beta$ are constants, and we obtain the solution
\begin{equation}
\begin{split}
&A(\theta_1,\theta_2)=C_1\theta_1\theta_2+\underline{C_2}\theta_2+\underline{C_3}\theta_1+C_4\\
&B(\theta_1,\theta_2)=C_5\theta_1\theta_2+\underline{C_6}\theta_2-\underline{C_3}\theta_1+C_7\\
&\Phi(t,\theta_1,\theta_2)=C_8\theta_1\theta_2+\underline{C_9}\theta_2+\gamma(t)\theta_1+\delta(t).
\end{split}
\label{solutionG1bis}
\end{equation}
This solution contains more freedom than (\ref{solutionG1}) in the constant parameters of the superfields $A$ and $B$, but less arbitrariness in the function $\Phi$. \\

\noindent In the case where $\underline{K_3}+\underline{K_7}\neq 0$, we get $\Phi_{t\theta_2}=0$ and $\Phi_{tt}=0$, and so, since $\Phi$ is of the form (\ref{formofphib612}), we obtain
\begin{equation}
\begin{split}
&A(\theta_1,\theta_2)=C_1\theta_1\theta_2+\underline{C_2}\theta_2+\underline{C_3}\theta_1+C_4\\
&B(\theta_1,\theta_2)=C_5\theta_1\theta_2+\underline{C_6}\theta_2+\underline{C_7}\theta_1+C_8\\
&\Phi(t,\theta_1,\theta_2)=C_9\theta_1\theta_2+\underline{C_{10}}\theta_2+C_{11}t\theta_1+C_{12}\theta_1+C_{13}t+C_{14}.
\end{split}
\label{solutionG1bisbis}
\end{equation}
Here, the function $\Phi$ is linear in $\theta_1$, $\theta_2$, $\theta_1\theta_2$ and $t$, and the superfields $A$ and $B$ are linear in $\theta_1$, $\theta_2$ and $\theta_1\theta_2$.\\

\noindent{\bf 2.} For the subalgebra ${\mathcal L}_2=\{\partial_t\}$, the invariants are $x$, $\theta_1$, $\theta_2$, $A$, $B$ and $\Phi$, which leads to the group orbit $A=A(x,\theta_1,\theta_2)$, $B=B(x,\theta_1,\theta_2)$, $\Phi=\Phi(x,\theta_1,\theta_2)$. Substituting into equations (\ref{eqmotion1B}), (\ref{eqmotion2B}) and (\ref{eqmotion3B}), we obtain seven different classes of solutions. The first solution is determined by three arbitrary functions, $A$, $B$ and $\Phi$, of the two fermionic variables $\theta_1$ and $\theta_2$:
\begin{equation}
A=A(\theta_1,\theta_2),\qquad
B=B(\theta_1,\theta_2),\qquad \Phi=\Phi(\theta_1,\theta_2).
\label{solutionG2no1}
\end{equation}
Next, we obtain the second solution in the form
\begin{equation}
\begin{split}
&A(x,\theta_1,\theta_2)=\theta_1\theta_2\alpha(x)+\theta_2c(x)+\theta_1b(x)+a(x)\\
&B(x,\theta_1,\theta_2)=C_1\theta_1\theta_2-\theta_1\theta_2\alpha(x)+\underline{C_2}\theta_2-\theta_2c(x)+\underline{C_3}\theta_1-\theta_1b(x)+C_4-a(x)\\
&\Phi(\theta_1,\theta_2)=\underline{C_5}\theta_2+\underline{C_6}\theta_1+C_7.
\end{split}
\label{solutionG2no2}
\end{equation}
where the superfields $A$ and $B$ depend on three arbitrary functions, $a$, $b$ and $c$, of $x$, and $\Phi$ depends linearly on $\theta_1$ and $\theta_2$. 
We then obtain the third solution in the rational form
\begin{equation}
\begin{split}
&A(x,\theta_1,\theta_2)=C_1+\underline{C_2}\theta_1+\underline{C_3}\theta_2+\theta_1\theta_2\alpha(x)\\
&B(x,\theta_1,\theta_2)=C_4+\underline{C_5}\theta_1+\underline{C_6}\theta_2-\theta_1\theta_2\alpha(x)\\
&\Phi(x,\theta_1,\theta_2)=C_7+\underline{C_8}\theta_1+\underline{C_9}\theta_2+\dfrac{\theta_1\theta_2K_0}{\alpha(x)}.
\end{split}
\label{solutionG2no3}
\end{equation}
which contains an arbitrary function $\alpha(x)$ in all three superfields, $A$, $B$ and $\Phi$.

\noindent The fourth solution
\begin{equation}
\begin{split}
&A(x,\theta_1,\theta_2)=C_1+\theta_1b(x)+\underline{C_2}\theta_2,\\
&B(x,\theta_1,\theta_2)=C_3+\underline{C_4}\theta_1-\theta_1b(x)+\underline{C_5}\theta_2,\\
&\Phi(x,\theta_1,\theta_2)=p(x)+\theta_1q(x)+\theta_2s(x).
\end{split}
\label{solutionG2no4}
\end{equation}
contains four arbitrary functions $b$, $p$, $q$ and $s$ of $x$.

\noindent The fifth type of solution is given by
\begin{equation}
\begin{split}
&A(x,\theta_1,\theta_2)=a(x)+\theta_1b(x)+\theta_2c(x)\\
&B(x,\theta_1,\theta_2)=C_1-a(x)+\underline{C_2}\theta_1-\theta_1b(x)+\underline{C_3}\theta_2-\theta_2c(x)\\
&\Phi(x,\theta_1,\theta_2)=C_4+\theta_1q(x)+\underline{C_5}\theta_2.
\end{split}
\label{solutionG2no5}
\end{equation}
where the superfields $A$ and $B$ depend on three arbitrary functions, $a$, $b$ and $c$, of $x$. 
For the sixth solution, $A$ and $B$ are linear in $\theta_1$ and $\theta_2$:
\begin{equation}
\begin{split}
&A(\theta_1,\theta_2)=C_1+\underline{C_2}\theta_1+\underline{C_3}\theta_2\\
&B(\theta_1,\theta_2)=C_4+\underline{C_5}\theta_1+\underline{C_6}\theta_2\\
&\Phi=\Phi(x,\theta_1,\theta_2).
\end{split}
\label{solutionG2no6}
\end{equation}
and $\Phi$ is an arbitrary function of one bosonic variable $x$ and two fermionic variables  $\theta_1$ and $\theta_2)$. 
Here, $A$ and $B$ are linear functions of $\theta_1$ and $\theta_2$. 
Finally, the seventh solution of the SUSY two-phase system (\ref{eqmotion1B}), (\ref{eqmotion2B}) and (\ref{eqmotion3B})
\begin{equation}
\begin{split}
&A(\theta_1,\theta_2)=C_1+\underline{C_2}\theta_1+\underline{C_3}\theta_2,\\
&B(x,\theta_1,\theta_2)=C_4-\varepsilon C_5 p(x)+\underline{C_6}\theta_1+\underline{C_7}\theta_2+C_5\theta_1\theta_2,\\
&\Phi(x,\theta_1,\theta_2)=p(x)+\underline{C_8}\theta_1+\underline{C_9}\theta_2+\varepsilon\theta_1\theta_2,
\end{split}
\label{solutionG2no7}
\end{equation}
contains one arbitrary function $p$ of $x$, which is present in both the superfields $B$ and $\Phi$, while $A$ depends on $\theta_1$ and $\theta_2$ only. \\

\noindent{\bf 3.} For the subalgebra ${\mathcal L}_3=\{\partial_x+a\partial_t\}$, the invariants are $\xi=ax-t$ (a symmetry variable), $\theta_1$, $\theta_2$, $A$, $B$ and $\Phi$, which leads to the group orbit $A=A(\xi,\theta_1,\theta_2)$, $B=B(\xi,\theta_1,\theta_2)$, $\Phi=\Phi(\xi,\theta_1,\theta_2)$. Substituting into the SUSY two-phase fluid flow equations (\ref{eqmotion1B}), (\ref{eqmotion2B}) and (\ref{eqmotion3B}), we obtain four different classes of solutions. First, we have
\begin{equation}
\begin{split}
&A(\theta_1,\theta_2)=A_0+\underline{A_1}\theta_1+\underline{A_2}\theta_2,\\
&B(\theta_1,\theta_2)=B_0+\underline{B_1}\theta_1+\underline{B_2}\theta_2,\\
&\Phi(x,t,\theta_1,\theta_2)=C_1(ax-t)+C_2+\underline{\Phi_1}\theta_1+\underline{\Phi_2}\theta_2+\Phi_3\theta_1\theta_2,
\end{split}
\label{solutionG3no1}
\end{equation}
where $\Phi$ is a simple travelling wave also involving the fermionic variables $\theta_1$ and $\theta_2$, while $A$ and $B$ depend linearly on $\theta_1$ and $\theta_2$. We obtain the solution
\begin{equation}
\begin{split}
&A(\theta_1,\theta_2)=A_0+\underline{A_1}\theta_1+\underline{A_2}\theta_2,\\
&B(\theta_1,\theta_2)=B_0-\underline{A_1}\theta_1+\underline{B_2}\theta_2,\\
&\Phi(x,t,\theta_1,\theta_2)=\phi_0(ax-t)+\underline{\Phi_1}\theta_1+\underline{\Phi_2}\theta_2+\Phi_3\theta_1\theta_2.
\end{split}
\label{solutionG3no2}
\end{equation}
which contains a travelling wave of arbitrary shape since it contains an arbitrary function $\phi_0$ of $ax-t$.

\noindent The solution
\begin{equation}
\begin{split}
&A(x,t,\theta_1,\theta_2)=(ax-t)^3+\underline{A_1}\theta_1+\underline{A_2}\theta_2,\\
&B(x,t,\theta_1,\theta_2)=-(ax-t)^3-\underline{A_1}\theta_1+\underline{B_2}\theta_2,\\
&\Phi(x,t,\theta_1,\theta_2)=-\dfrac{1}{ax-t}+\underline{\phi_1}\theta_1+\underline{\phi_2}\theta_2+\dfrac{1}{a}\theta_1\theta_2.
\end{split}
\label{solutionG3no3}
\end{equation}
represents a travelling wave which is algebraic in $(ax-t)$. Finally, the solution
\begin{equation}
\begin{split}
&A(x,t,\theta_1,\theta_2)=\dfrac{K_0}{C_2}(ax-t)+C_4+\underline{A_1}\theta_1+\underline{A_2}\theta_2,\\
&B(x,t,\theta_1,\theta_2)=-\dfrac{K_0}{C_2}(ax-t)-C_4+C_1+\underline{B_1}\theta_1+\underline{B_2}\theta_2,\\
&\Phi(x,t,\theta_1,\theta_2)=C_2(ax-t)+C_3+\underline{\phi_1}\theta_1+\underline{\phi_2}\theta_2+\dfrac{1}{a}\theta_1\theta_2,
\end{split}
\label{solutionG3no4}
\end{equation}
contains a linear travelling wave in all three superfields $A$, $B$ and $\Phi$, which is linear in $(ax-t)$.\\

\noindent{\bf 4.} For the subalgebra ${\mathcal L}_5=\{(1-\underline{\mu}\theta_1)\partial_x+\mu\partial_{\theta_1}\}$, the invariants are $t$, $\eta_1=\theta_1-\underline{\mu}x$, $\theta_2$, $A$, $B$ and $\Phi$, which leads to the group orbit $A=A(t,\eta_1,\theta_2)$, $B=B(t,\eta_1,\theta_2)$, $\Phi=\Phi(t,\eta_1,\theta_2)$. Substituting into the SUSY two-phase equations (\ref{eqmotion1B}), (\ref{eqmotion2B}) and (\ref{eqmotion3B}), we find the following solution:
\begin{equation}
\begin{split}
&A(x,t,\theta_1,\theta_2)=-\underline{\mu}A_3\underline{\phi_1}t+\underline{A_1}(\theta_1-\underline{\mu}x)+\underline{A_2}\theta_2+A_3(\theta_1-\underline{\mu}x)\theta_2+C_1,\\
&B(x,t,\theta_1,\theta_2)=\underline{\mu}A_3\underline{\phi_1}t-\underline{A_1}(\theta_1-\underline{\mu}x)+\underline{B_2}\theta_2-A_3(\theta_1-\underline{\mu}x)\theta_2+C_2,\\
&\Phi(x,t,\theta_1,\theta_2)=\phi_0(t)+\underline{\phi_1}(\theta_1-\underline{\mu}x)+\underline{\phi_2}\theta_2,
\end{split}
\label{solutionG4no1}
\end{equation}
where all three superfields depend linearly on $\theta_1-\underline{\mu}x$, while $\Phi$ contains an arbitrary function of $\phi_0$ of $t$.\\

\noindent{\bf 5.} For the subalgebra ${\mathcal L}_5=\{\partial_x-\underline{\mu}\theta_2\partial_t+\mu\partial_{\theta_2}\}$, the invariants are $\xi=t+\underline{\mu}\theta_2x$ (the symmetry variable), $\theta_1$, $\eta_2=\theta_2-\underline{\mu}x$, $A$, $B$ and $\Phi$, which leads to the group orbit $A=A(\xi,\theta_1,\eta_2)$, $B=B(\xi,\theta_1,\eta_2)$, $\Phi=\Phi(\xi,\theta_1,\eta_2)$. Substituting into the SUSY equations (\ref{eqmotion1B}), (\ref{eqmotion2B}) and (\ref{eqmotion3B}), we get the solution:
\begin{equation}
\begin{split}
&A(x,t,\theta_1,\theta_2)=-\underline{\mu}\underline{A_2}\phi_3(t+\underline{\mu}\theta_2x)+\underline{A_1}\theta_1+\underline{A_2}(\theta_2-\underline{\mu}x),\\
&B(x,t,\theta_1,\theta_2)=\underline{\mu}\underline{A_2}\phi_3(t+\underline{\mu}\theta_2x)-\underline{A_1}\theta_1-\underline{A_2}(\theta_2-\underline{\mu}x),\\
&\Phi(x,t,\theta_1,\theta_2)=\phi_0(t+\underline{\mu}\theta_2x)+\underline{\phi_1}\theta_1+\underline{\phi_2}(\theta_2-\underline{\mu}x)+\phi_3\theta_1(\theta_2-\underline{\mu}x),
\end{split}
\label{solutionG5no1}
\end{equation}
which contains an arbitrary function $\phi_0$ of $(t+\underline{\mu}\theta_2 x)$ and has a linear dependence on $(\theta_2-\underline{\mu}x)$. \\

\noindent{\bf 6.} For the subalgebra ${\mathcal L}_{16}=\{2x\partial_x+2t\partial_t+\theta_1\partial_{\theta_1}+\theta_2\partial_{\theta_2}+2\Phi\partial_{\Phi}\}$, the invariants are $\xi=\dfrac{x}{t}$ (a symmetry variable), $\eta_1=\dfrac{\theta_1}{\sqrt{t}}$, $\eta_2=\dfrac{\theta_2}{\sqrt{t}}$, $A$, $B$ and $F=\dfrac{\Phi}{t}$, which leads to the group orbit $A=A(\xi,\eta_1,\eta_2)$, $B=B(\xi,\eta_1,\eta_2)$, $\Phi=tF(\xi,\eta_1,\eta_2)$. Substituting into equations (\ref{eqmotion1B}), (\ref{eqmotion2B}) and (\ref{eqmotion3B}), we obtain the invariant solution
\begin{equation}
\begin{split}
&A(x,t,\theta_1,\theta_2)=2A_3\sqrt{\dfrac{x-F_3t}{t}}+\dfrac{A_3}{t}\theta_1\theta_2\\
&B(x,t,\theta_1,\theta_2)=-2A_3\sqrt{\dfrac{x-F_3t}{t}}-\dfrac{A_3}{t}\theta_1\theta_2\\
&\Phi(x,t,\theta_1,\theta_2)=\dfrac{2t}{3}\left(\dfrac{x-F_3t}{t}\right)^{3/2}+F_3\theta_1\theta_2.
\end{split}
\label{solutionG5no1}
\end{equation}
which has a dependence on the quantity $\frac{x-F_3t}{t}$, where $A_3$ and $F_3$ are arbitrary constants. The solution (\ref{solutionG5no1}) represents a center wave in the sense proposed in classical gas dynamics \cite{Roz,Sidorov}.

\section{Concluding Remarks}
\label{sec:5}

\noindent In this paper we discuss the construction of a supersymmetric extension of the two-phase fluid flow system (\ref{eq1.4}) through a superspace and superfield formalism. This analysis includes a supersymmetric extension of a one-dimensional ideal compressible non-viscous two-phase fluid flow. This allows us to determine a Lie superalgebra of infinitesimal symmetries which generate the Lie point symmetries of this system of equations. We observe that, in analogy with a classical Euler system written in terms of Riemann invariants, the symmetry superalgebra of the supersymmetric two-phase fluid flow system (\ref{eqmotion1B}), (\ref{eqmotion2B}) and (\ref{eqmotion3B}) contains two independent dilations. Next, we discuss the classification of their subalgebras. 
A systematic classification in terms of conjugacy classes was performed for the one-dimensional subalgebras, resulting in a list of 63 nonequivalent classes of subalgebras. Consequently, a complete symmetry reduction analysis of this supersymmetric fluid system would lead to very large classes of invariant solutions. Through the use of a generalized version of the symmetry reduction method we have demonstrated how to find exact invariant solutions of the supersymmetric system. A systematic use of the structure of the invariance supergroups of the two phase fluid flow equations allows us to generate symmetry variables. For certain subalgebras, the invariants have a nonstandard structure and therefore do not lead to invariant solutions. The symmetry reduction method allows us to reduce, after some transformations, the initial system of PDEs to many possible ODEs. We show that in each case, the three superfields, $A$, $B$ and $\Phi$, can be decomposed in terms of their independent fermionic variables, $\theta_1$ and $\theta_2$, and their bosonic combination $\theta_1\theta_2$. This allows us to determine the invariant solutions of the supersymmetric two-phase fluid flow system component-wise. For the purpose of illustration, for six specific subalgebras involving different types of generators, a number of invariant solutions were found. These solutions involve several arbitrary functions. The most general solutions are expressed in terms of one or two fermionic and one bosonic variables. Such a large amount of arbitrariness  was not found in previous analyses by the authors of supersymmetric hydrodynamic-type systems (i.e. the Euler system \cite{EulerSUSY}, polytropic gas \cite{Polytropic}, systems written in terms of Riemann invariants \cite{GHhydroRiemann} and Gaussian fluid flow \cite{Gaussian}. Some of the obtained solutions involve damping and growth. However, in contrast with their classical counterparts, no blow-up phenomenon (i.e. the gradient catastrophe) was observed for any solution presented here.

\noindent On the basis of the considerations presented in this paper, the question of the solvability of the Cauchy problem for supersymmetric versions of hydrodynamic-type systems (involving the continuous dependence on the initial data) arises. More precisely, given some Cauchy data at $t=t_0$, does it satisfy the conditions for the formulation and mathematical correctness of problems involving the evolution in time $t$ of multiwave solutions admitted by the SUSY system under consideration. This analysis could be done, for instance, for the examples considered here. From the known analytical dependence of a group-invariant solution, one can attempt to find the functional dependence of the Cauchy data. This approach would determine the arbitrary functions appearing in invariant solutions. This subject will be addressed in a future work.

\section{Appendix}
\label{sec:6}

The following list constitutes the classification of the one-dimensional subalgebras of the Lie symmetry superalgebra associated with the supersymmetric system (\ref{eqmotion1B}), (\ref{eqmotion2B}) and (\ref{eqmotion3B}) into conjugacy classes under the action of the associated Lie group. Here $\varepsilon=\pm 1$, the parameters $a$ and $b$ are non-zero bosonic constants, and $\underline{\mu}$ and $\underline{\nu}$ are non-zero fermionic constants.

\begin{equation}
\begin{split}
&{\mathcal L}_1=\{P_1\}, \hspace{5mm} {\mathcal L}_2=\{P_2\}, \hspace{5mm}
{\mathcal L}_3=\{P_1+aP_2\}, \hspace{5mm} {\mathcal L}_4=\{\underline{\mu}Q_1\}, \\ & {\mathcal L}_5=\{P_1+\underline{\mu}Q_1\}, \hspace{5mm}
{\mathcal L}_6=\{P_2+\underline{\mu}Q_1\}, \hspace{5mm} {\mathcal L}_7=\{P_1+aP_2+\underline{\mu}Q_1\}, \\ & {\mathcal L}_8=\{\underline{\mu}Q_2\}, \hspace{5mm} {\mathcal L}_9=\{P_1+\underline{\mu}Q_2\}, \hspace{5mm} {\mathcal L}_{10}=\{P_2+\underline{\mu}Q_2\}, \\ &
{\mathcal L}_{11}=\{P_1+aP_2+\underline{\mu}Q_2\}, \hspace{5mm} {\mathcal L}_{12}=\{\underline{\mu}Q_1+\underline{\nu}Q_2\}, \hspace{5mm}
{\mathcal L}_{13}=\{P_1+\underline{\mu}Q_1+\underline{\nu}Q_2\}, \\ & {\mathcal L}_{14}=\{P_2+\underline{\mu}Q_1+\underline{\nu}Q_2\}, \hspace{5mm} {\mathcal L}_{15}=\{P_1+aP_2+\underline{\mu}Q_1+\underline{\nu}Q_2\}, \hspace{5mm} {\mathcal L}_{16}=\{M_1\}, \\ & 
{\mathcal L}_{17}=\{M_1+\varepsilon P_1\}, \hspace{5mm} {\mathcal L}_{18}=\{M_1+\varepsilon P_2\},  \hspace{5mm}
{\mathcal L}_{19}=\{M_1+\varepsilon P_1+aP_2\}, \\ & {\mathcal L}_{20}=\{M_1+\underline{\mu}Q_1\},  \hspace{5mm}
{\mathcal L}_{21}=\{M_1+\varepsilon P_1+\underline{\mu}Q_1\}, \hspace{5mm} {\mathcal L}_{22}=\{M_1+\varepsilon P_2+\underline{\mu}Q_1\},  \\ &
{\mathcal L}_{23}=\{M_1+\varepsilon P_1+aP_2+\underline{\mu}Q_1\}, \hspace{5mm} {\mathcal L}_{24}=\{M_1+\underline{\mu}Q_2\},  \\ &
{\mathcal L}_{25}=\{M_1+\varepsilon P_1+\underline{\mu}Q_2\}, \hspace{5mm} {\mathcal L}_{26}=\{M_1+\varepsilon P_2+\underline{\mu}Q_2\},  \\ &
{\mathcal L}_{27}=\{M_1+\varepsilon P_1+aP_2+\underline{\mu}Q_2\}, \hspace{5mm} {\mathcal L}_{28}=\{M_1+\underline{\mu}Q_1+\underline{\nu}Q_2\},  \\ &
{\mathcal L}_{29}=\{M_1+\varepsilon P_1+\underline{\mu}Q_1+\underline{\nu}Q_2\}, \hspace{5mm} {\mathcal L}_{30}=\{M_1+\varepsilon P_2+\underline{\mu}Q_1+\underline{\nu}Q_2\}, \\ & {\mathcal L}_{31}=\{M_1+\varepsilon P_1+aP_2+\underline{\mu}Q_1+\underline{\nu}Q_2\}, \hspace{5mm} {\mathcal L}_{32}=\{M_2\}, \hspace{5mm} {\mathcal L}_{33}=\{M_2+\varepsilon P_1\}, \\ & {\mathcal L}_{34}=\{M_2+\varepsilon P_2\}, \hspace{5mm}
{\mathcal L}_{35}=\{M_2+\varepsilon P_1+aP_2\}, \hspace{5mm} {\mathcal L}_{36}=\{M_2+\underline{\mu}Q_1\}, \\ & {\mathcal L}_{37}=\{M_2+\varepsilon P_1+\underline{\mu}Q_1\}, \hspace{5mm}
{\mathcal L}_{38}=\{M_2+\varepsilon P_2+\underline{\mu}Q_1\}, \\ & {\mathcal L}_{39}=\{M_2+\varepsilon P_1+aP_2+\underline{\mu}Q_1\}, \hspace{5mm} {\mathcal L}_{40}=\{M_2+\underline{\mu}Q_2\}, \\ & {\mathcal L}_{41}=\{M_2+\varepsilon P_1+\underline{\mu}Q_2\}, \hspace{5mm} {\mathcal L}_{42}=\{M_2+\varepsilon P_2+\underline{\mu}Q_2\}, \\ &
{\mathcal L}_{43}=\{M_2+\varepsilon P_1+aP_2+\underline{\mu}Q_2\}, \hspace{5mm} {\mathcal L}_{44}=\{M_2+\underline{\mu}Q_1+\underline{\nu}Q_2\}, \\ &
{\mathcal L}_{45}=\{M_2+\varepsilon P_1+\underline{\mu}Q_1+\underline{\nu}Q_2\}, \hspace{5mm} {\mathcal L}_{46}=\{M_2+\varepsilon P_2+\underline{\mu}Q_1+\underline{\nu}Q_2\}, \\ & {\mathcal L}_{47}=\{M_2+\varepsilon P_1+aP_2+\underline{\mu}Q_1+\underline{\nu}Q_2\}, \hspace{5mm} {\mathcal L}_{48}=\{M_2+aM_1\}, \\ &
{\mathcal L}_{49}=\{M_2+aM_1+\varepsilon P_1\}, \hspace{5mm} {\mathcal L}_{50}=\{M_2+aM_1+\varepsilon P_2\},  \\ &
{\mathcal L}_{51}=\{M_2+aM_1+\varepsilon P_1+bP_2\}, \hspace{5mm} {\mathcal L}_{52}=\{M_2+aM_1+\underline{\mu}Q_1\},  \\ &
{\mathcal L}_{53}=\{M_2+aM_1+\varepsilon P_1+\underline{\mu}Q_1\}, \hspace{5mm} {\mathcal L}_{54}=\{M_2+aM_1+\varepsilon P_2+\underline{\mu}Q_1\},  \\ &
{\mathcal L}_{55}=\{M_2+aM_1+\varepsilon P_1+bP_2+\underline{\mu}Q_1\}, \hspace{5mm} {\mathcal L}_{56}=\{M_2+aM_1+\underline{\mu}Q_2\},  \\ &
{\mathcal L}_{57}=\{M_2+aM_1+\varepsilon P_1+\underline{\mu}Q_2\}, \hspace{5mm} {\mathcal L}_{58}=\{M_2+aM_1+\varepsilon P_2+\underline{\mu}Q_2\},  \\ &
{\mathcal L}_{59}=\{M_2+aM_1+\varepsilon P_1+bP_2+\underline{\mu}Q_2\}, \hspace{5mm} {\mathcal L}_{60}=\{M_2+aM_1+\underline{\mu}Q_1+\underline{\nu}Q_2\}, \\ & {\mathcal L}_{61}=\{M_2+aM_1+\varepsilon P_1+\underline{\mu}Q_1+\underline{\nu}Q_2\}, \hspace{5mm} {\mathcal L}_{62}=\{M_2+aM_1+\varepsilon P_2+\underline{\mu}Q_1+\underline{\nu}Q_2\}, \\ & {\mathcal L}_{63}=\{M_2+aM_1+\varepsilon P_1+bP_2+\underline{\mu}Q_1+\underline{\nu}Q_2\}
\end{split}
\label{list5}
\end{equation}


%
%

\begin{acknowledgements}
Both authors thank the Mathematical Physics Laboratory of the Centre de Recherches Math\'ematiques, Universit\'e de Montr\'eal for its support during the writing of this paper.
\end{acknowledgements}

%
 \section*{Conflict of interest}

 The authors declare that they have no conflict of interest.


\begin{thebibliography}{}
%
%

\bibitem{CourantHilbert}
Courant, R., Hilbert, D.: Methods of Mathematical Physics, Vol. 2. Interscience, New York (1962)

\bibitem{Mises}
Mises, R.: Mathematical Theory of Compressible Fluid Flow. Academic Press, New York (1958)

\bibitem{Lighthill}
Lighthill, H.: Hyperbolic Equations and Waves. Springer-Verlag, New York (1968)

\bibitem{Whitham}
Whitham, G.B.: Linear and Nonlinear Waves. John Wiley Pub., New York (1974)

\bibitem{Lamb}
Lamb, H.: Hydrodynamics. Cambridge Univ. Press, Cambridge (1993)

\bibitem{Jeffrey}
Jeffrey, A.: Quasilinear Hyperbolic Systems and Waves. Pitman, San Francisco (1976)

\bibitem{Roz}
Rozdestvenskii, B., Janenko, N.: Systems of Quasilinear Equations and Their Applications to Gas Dynamics. Translation Math. Monographs, Vol. 55. AMS, Providence, RI (1983)

\bibitem{Godunov}
Godunov, S.K.: Mathematical Physics Equations (in Russian). Nauka, Moscow (1979)

\bibitem{Jackiw}
Jackiw, R.: A Particle Theorist's Lectures on Supersymmetric Non-Abelian Fluid Mechanics and d-Branes. Springer, New York (2002)

\bibitem{Majda}
Majda, A.: Compressible Fluid Flow and Systems of Conservation Laws in Several Space Variables. Springer-Verlag, New York (1984)

\bibitem{Peradzynski}
Peradzynski, Z.: Advances in Nonlinear Waves. Research Notes in Math. Vol. 111, Ed. L. Debnath Pitman, Boston (1985)

\bibitem{Ovsiannikov}
Ovsiannikov, L.V.: Group Analysis of Differential Equations. Academic Press, New York (1982)

\bibitem{Zakharov}
Zakharov, V.: Nonlinear Waves and Weak Turbulence. Advances of Modern Mathematics Translation, Series 2 AMS, Providence, RI (1997)

\bibitem{Riemann1}
Riemann, B.: Versuch Einer Allgemeinen Auffassung der Integration und Differentiation. Teubner, Leipzig, (1876); ibid. Dover, New York (1953), pp 331--344

\bibitem{Riemann2}
Riemann, B.: \"Uber die Fortpflanzung Ebener Luftwellen von Endlicher Schwingungsweite. Teubner, Leipzig, (1876); ibid. Dover, New York (1953), pp 145--164

\bibitem{Poisson}
Poisson, S.D.: M\'emoire sur la th\'eorie du son.
Journal de l'\'Ecole Polytechnique, 14$^{\mbox{i\`eme}}$ cahier, 7, Paris pp 319--392 (1808)

\bibitem{CourantFriedrichs}
Courant, R. , Friedrichs, K.O.: Supersonic Flow and Shock Waves. Interscience, New York (1948)

\bibitem{Hugoniot}
Hugoniot, H.:
J. Ec. Polytech. (Paris) \textbf{1}, 57--79 (1887)

\bibitem{Opanasenko1}
Opanasenko, S., Bihlo, A., Popovych, R.O., Sergeyev, A.: Generalized symmetries, conservation laws and Hamiltonian structures of an isothermal no-slip drift flux model,
Phys. D. \textbf{411}, 132546 (2020)

\bibitem{Opanasenko2}
Opanasenko, S., Bihlo, A., Popovych, R.O., Sergeyev, A.: Extended symmetry analysis of isothermal no-slip drift flux model,
Phys. D. \textbf{402}, 132188 (2020)

\bibitem{Panov}
Panov, A.V.: Invariant solutions and submodels in two-phase fluid mechanics generated by 3-dimensional subalgebras, Barochronous flows,
Int. J. Non-Linear Mech. \textbf{116}, 140--146 (2019)

\bibitem{John1}
John, F.: Nonlinear Wave Equations, Formation of Singularities, University Lecture Series 2. AMS, Providence (1990)

\bibitem{John2}
John, F.: Formulation of singularities in one-dimensional nonlinear wave propagation,
Commun. Pure Appl. Math. \textbf{27}, 377--394 (1974)

\bibitem{Cornwell}
Cornwell, J. F.: Group Theory in Physics, Vol. 3. Academic Press, London (1989)

\bibitem{DeWitt}
DeWitt, B.: Supermanifolds. Cambridge University Press, Cambridge (1984)

\bibitem{Rogers1}
Rogers, A.: A global theory of Supermanifolds,
J. Math. Phys. \textbf{21}, 1352--1365 (1980)

\bibitem{Rogers2}
Rogers, A.: Supermanifolds: Theory and Applications. World Scientific, London (2007)

\bibitem{Manin}
Manin, Y.I., Radul, A.O.: (the title),
Commun. Math. Phys. \textbf{98}, 65 (1985)

\bibitem{Fatyga}
Fatyga, B.W., Kostelecky, V.A., Truax, D.R.: Grassmann-valued fluid dynamics,
J. Math. Phys. \textbf{30}, 1464--1472 (1989)

\bibitem{EulerSUSY}
Grundland, A.M., Hariton, A.J.: Supersymmetric version of the Euler system and its invariant solutions,
Symmetry. \textbf{5}, 253--270 (2013)

\bibitem{Polytropic}
Grundland, A.M., Hariton, A.J.: Supersymmetric formulation of polytropic gas dynamics and its invariant solutions,
J. Math. Phys. \textbf{52}, 043501 (2011)

\bibitem{GHhydroRiemann}
Grundland, A.M., Hariton, A.J.: Supersymmetric version of a hydrodynamic system in Riemann invariants and its solutions,
J. Math. Phys. \textbf{49}, 043502 (2008)

\bibitem{Gaussian}
Grundland, A.M., Hariton, A.J.: Supersymmetric version of a Gaussian irrotational compressible fluid flow,
J. Phys. A: Math. Theor. \textbf{40}, 15113--15129 (2007)

\bibitem{Freed}
Freed, D.S.: Five Lectures on Supersymmetry. AMS, Providence, RI (1999)

\bibitem{Varadarajan}
Varadarajan, V.S.: Reflections on Quanta, Symmetries and Supersymmetries. Springer, New York (2011)

\bibitem{Olver}
Olver, P.J.: Applications of Lie Groups to Differential Equations. Springer, New York (1986)

\bibitem{Sidorov}
Sidorov, A.F., Shapeev, V., Janenko, N.: Method of Differential Constraints Applied to Gas Dynamics (in Russian). Nauka, Moscow (1984)

\bibitem{Winternitz}
Winternitz, P.: Lie Groups and Solutions of Nonlinear Partial Differential Equations, in Integrable Systems, Quantum Groups and Quantum Field Theories. Eds. L.A. Ibort and M.A. Rodriguez, p. 429. Kluwer, Dordrecht (1993)

\bibitem{Goursat1}
Goursat, E.: Sur les substitutions orthogonales et les divisions r\'{e}guli\`{e}res de l'espace,
Ann. Sci. Ec. Norm. Sup. \textbf{6}, 3, 9--102 (1880)

\bibitem{Clarkson}
Clarkson, P.A., Winternitz, P.: Symmetry reduction and exact solutions of nonlinear partial differential equations, in The Painlev\'e Property, One Century Later. Ed. R. Conte, p. 597. Springer-Verlag, New York (1999)

\end{thebibliography}


\end{document}